\documentclass[draftcls,onecolumn,letterpaper]{IEEEtran}
\usepackage{times,amsmath,epsfig,amssymb}
\usepackage{epstopdf}
\usepackage{caption,cite,textcomp}
\usepackage{graphicx,subfigure,url}
\begin{document}
\title{Reconfigurable nanoelectronics using graphene based spintronic logic gates}

%>>>> The author is responsible for formatting the
%  author list and their institutions.  Use  \skiplinehalf
%  to separate author list from addresses and between each address.
%  The correspondence between each author and his/her address
%  can be indicated with a superscript in italics,
%  which is easily obtained with \supit{}.

\author{Hanan~Dery,~\IEEEmembership{Member,~IEEE,}~Hui~Wu,~\IEEEmembership{Member,~IEEE,}~Berkehan Ciftcioglu,~\IEEEmembership{Member,~IEEE,}~ Michael~Huang,~\IEEEmembership{Member,~IEEE,}~Yang~Song, Roland~Kawakami, Jing~Shi, Ilya~Krivorotov,~\IEEEmembership{Member,~IEEE,}~Donald~A.~Telesca,~Igor~\v{Z}uti\'{c},~and Lu~J.~Sham
\thanks{Hanan Dery, Hui Wu, Berkehan Ciftcioglu, and Michael Huang are with the Department of Electrical and Computer Engineering, University of Rochester, Rochester, NY 14627. Yang Song is with the Department of Physics, University of Rochester, Rochester, NY 14627. Roland Kawakami and Jing Shi are with the Department of Physics,  University of California Riverside, Riverside, CA 92521. Ilya Krivorotov is with the Department of Physics,  University of California Irvine, Irvine, CA 92697.  Donald A. Telesca is with the Space Electronics Branch, Air Force Research Lab (AFRL/RVSEF), Kirtland AFB, NM  87117. Igor \v{Z}uti\'{c} is with the Department of Physics,  State University of New-York at Buffalo, Buffalo, NY 14260. Lu J. Sham is with the Department of Physics, University of California San Diego, San
Diego, CA 92093.} \thanks{This work is supported by AFOSR Contract No. FA9550-09-1-0493 and NSF Contract No. ECCS-0824075.}
\thanks{Email: hanan.dery@rochester.edu}}

%
%\date{}
\maketitle

%>>>> Further information about the authors, other than their
%  institution and addresses, should be included as a footnote,
%  which is facilitated by the \authorinfo{} command.

%\authorinfo{Further author information: (Send correspondence to H.D.)\\H.D.: E-mail: hanan.dery@rochester.edu, Telephone: 1 585 275 3870}
%%>>>> when using amstex, you need to use @@ instead of @

%%%%%%%%%%%%%%%%%%%%%%%%%%%%%%%%%%%%%%%%%%%%%%%%%%%%%%%%%%%%%
%>>>> uncomment following for page numbers
% \pagestyle{plain}
%>>>> uncomment following to start page numbering at 301
%\setcounter{page}{301}

%%%%%%%%%%%%%%%%%%%%%%%%%%%%%%%%%%%%%%%%%%%%%%%%%%%%%%%%%%%%%
\begin{abstract}
This paper presents a novel design concept for spintronic nanoelectronics that emphasizes a seamless integration of spin-based memory and logic circuits. The building blocks are magneto-logic gates \cite{Dery_Nature07} based on a hybrid graphene/ferromagnet material system. We use network search engines as a technology demonstration vehicle and present a spin-based circuit design with smaller area, faster speed, and lower energy consumption than the state-of-the-art CMOS counterparts. This design can also be applied in applications such as data compression \cite{Kom93,Wei93,Yan94,Cra98}, coding \cite{Liu94} and image recognition \cite{Mer00,Nak00}. In the proposed scheme, over 100 spin-based logic operations are carried out before any need for a spin-charge conversion. Consequently, supporting CMOS electronics requires little power consumption. The spintronic-CMOS integrated system can be implemented on a single 3-D chip. These nonvolatile logic circuits hold potential for a paradigm shift in computing applications.
\end{abstract}

%>>>> Include a list of keywords after the abstract

%%%%%%%%%%%%%%%%%%%%%%%%%%%%%%%%%%%%%%%%%%%%%%%%%%%%%%%%%%%%%
\IEEEpeerreviewmaketitle


\section{Introduction}
\label{sec:intro}

\PARstart{T}{he} continued Moore's law scaling in CMOS integrated circuits poses increasing challenges to provide low-energy consumption, sufficient processor speed, bandwidth of interconnects, and memory storage \cite{ITRS09,Rakheja10}. CMOS logic circuits rely on the von Neumann computer architecture consisting of central processing units (CPU) connected by some communication channel to memory.  The bottleneck caused by the communication (sometimes dubbed as the von Neumann bottleneck) and memory accesses is the underlying reason for the significant and widening gap between the fast improving transistor performance and our relatively stagnant ability to provide correspondingly faster program executions. Such bottlenecks are especially obvious for data intensive applications where most of the actions involve accessing or checking data (rather than doing complex computation). Network routers are a classical example where the Internet protocol (IP) address is compared to a list of patterns to find the best match for further processing. In a typical router, special-purpose associative tables are used where the search data (the key) is simultaneously compared with all of the table entries to find matches.  Such tables allow us to fundamentally improve the processing speed. Unfortunately, conventional CMOS implementation of these circuits also suffer from scalability issues, making them ineffective for larger search problems that are increasingly relevant to modern workloads.

In this paper, we propose a paradigm change for these search applications
using spintronics, which posses fundamental advantages over CMOS in chip area
and power consumption. Specifically, we demonstrate a 3.2-Mbit spintronic
search engine with 25K words of 128 bits using 0.04-$\mu$m$^{2}$ bit
cells, designed and simulated using device and circuit simulation. In this
design, the total chip area is less than 1~mm$^{2}$. For the same die
size, Analog Bits, Inc. provides a macro 50x smaller (64Kb) using TSMC
0.13$\mu m$ technology. Also for comparison, the reported package size of a
state-of-the-art CMOS based search engine is 729 mm$^{2}$ (Renesas R8A20410BG;
20 Mbits). As important, the total power consumption of this spintronic search
engine without any design optimization is smaller than state-of-the-art CMOS
based search engines. The proposed spintronic circuit architecture can also
enhance a variety of applications that benefit from massive parallel search
operations (e.g.; data compression \cite{Kom93,Wei93,Yan94,Cra98}, coding
\cite{Liu94} and image recognition \cite{Mer00,Nak00}).

This paper is organized as follow. Section~\ref{sec:mlg} discusses the MLG basic device operation. It includes explanations on the materials choice and on spin properties. In Sec.~\ref{sec:circuit} we present the spintronic search engine circuit design and we discuss its efficient searching methods as well as some of its scaling merits. Section~\ref{sec:design} discusses a proposed spintronics-CMOS interface which is then simulated by an advanced design system. In this part we also discuss power dissipation aspects. In Sec.~\ref{sec:apps} we briefly mention other potential applications of the proposed spintronic system and in Sec.~\ref{sec:conc} we summarize our findings.

\section{The building block: Magneto-Logic Gate (MLG)}
\label{sec:mlg}

\begin{figure}
\center{\includegraphics[width=3.5in]{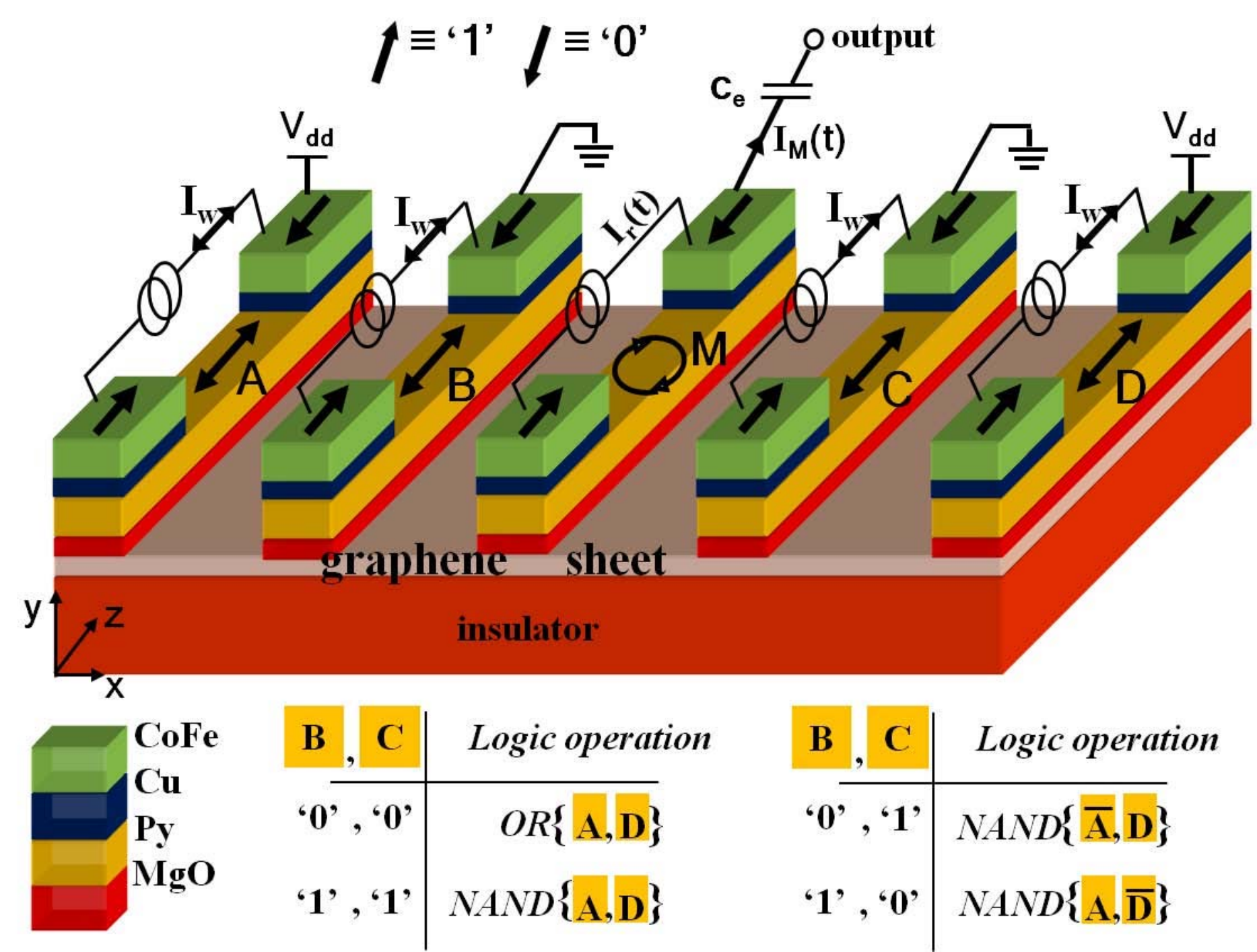}} \caption{A universal and reconfigurable magneto-logic gate (MLG). Five magnetic terminals are deposited on top of a single layer graphene sheet. The spin accumulation profile in the sheet determines the logic result and it is governed by the magnetic directions of the biased sections (A-B and C-D). Using spin-transfer torque, the logic operands (magnetization directions of A-D) are encoded via the individual writing currents, I$_w$(t), across the low resistive and
all-metallic path (CoFe/Cu/Py/Cu/CoFe). The readout is triggered by the reading current signal, I$_r$(t) that perturbs the magnetization of the middle contact. The binary logic output is the resulting on/off transient current I$_M$(t). \label{fig:mlg}}
\end{figure}

A \emph{magneto-logic gate} (MLG) is adopted as the basic building block due to its potentially favorable properties of spin amplification, speed and scalability \cite{Dery_Nature07}. Fig.~\ref{fig:mlg} shows a universal and reconfigurable MLG that consists of five ferromagnetic (FM) electrodes on top of a graphene layer. FM regions are inherently nonvolatile, they preserve the direction of magnetization even in the absence of any power supply. While this feature has been extensively used for robust information
storage in magnetic hard drives and magnetic random access memories (MRAMs) \cite{Zutic_RMP04}, here we show how this nonvolatility can also be used for high-performance magneto-logic. The magnetization itself reflects that the FM electrode has an unequal number of electrons with two different spin projections (up and down; minority and majority). The MLG design employs a stack of FM layers where the elongated permalloy layer (Py) is the free magnetic layer into which the information is encoded. Details of the FM layered system will be explained in Sec.~\ref{sec:STT}. The MLG operation relies on the generation of non-equilibrium spin accumulation when spin polarized electrons tunnel from the free layer into the graphene via the MgO tunneling barrier. The magnitude of the spin accumulation in the graphene strongly depends on the relative orientation of the
magnetization directions in the free layers of the MLG. Of the five FM contacts, the middle contact (M) is used for readout and the remaining four contacts (A, B, C \& D) are logic operands whose values are defined by the magnetization direction (`0' \& `1' in Fig.~\ref{fig:mlg}). The output state is given by the Boolean expression (A~xor~B) + (C~xor~D). By reprogramming the B and C states we get a universal set of four logic operations between A and D.

\begin{figure}
 \center{\includegraphics[width=2.3in]{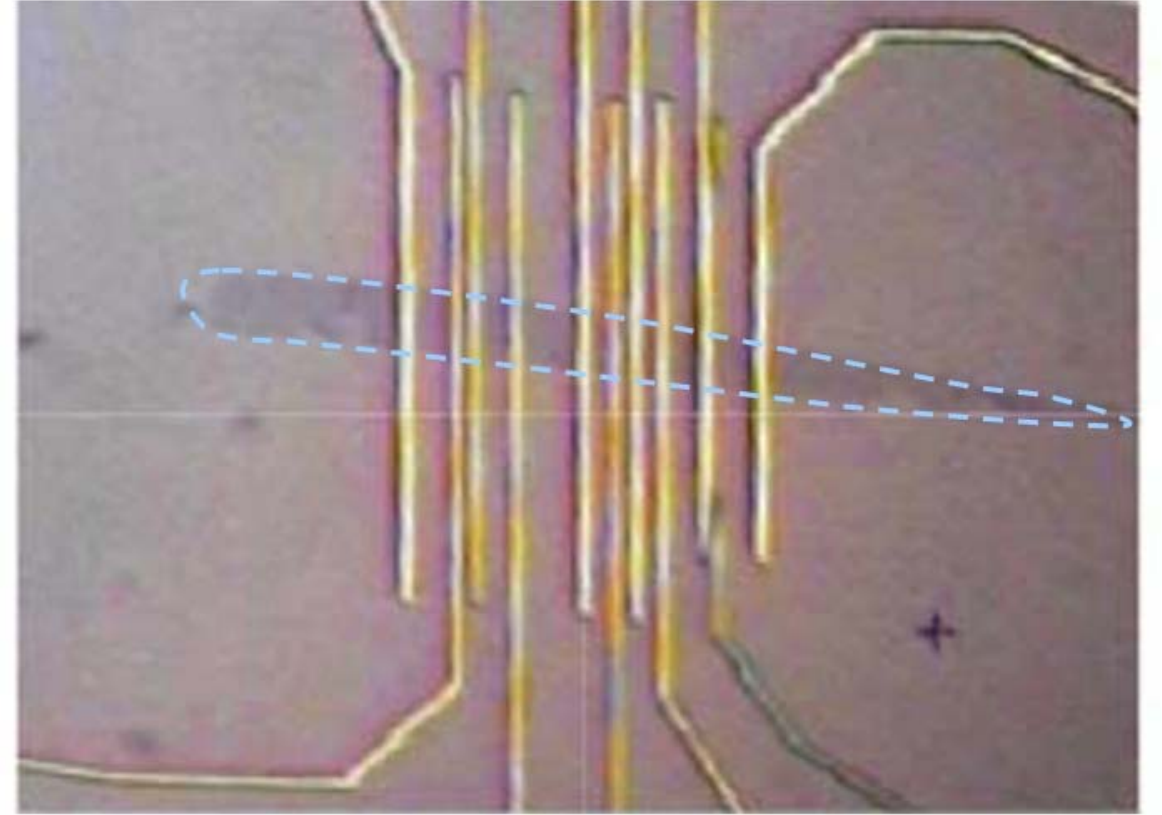}}
 \caption{An optical microscope image of a lateral graphene sheet topped with ultrathin Ti-seeded MgO tunnel barrier and Co contacts. The graphene flake is bounded by the dash line. This structure has been used to extract spin dependent graphene parameters in Ref.~\cite{Han10}.
 \label{fig:image}}
\end{figure}

The logic operation is triggered by perturbing the magnetization direction of M. The electrical response to the perturbation is governed by the potential level in the middle contact. In steady-state, the potential level is set by the zero electrical current condition, $I_m(t)$=0, due to the external capacitor, $C_e$. In the case of small external and intrinsic device capacitance, the response is instantaneous and the potential level `follows' the magnetization direction. The transient current response depends on the $RC$ of the system and on the spin-accumulation profile in the graphene layer. This profile is a function of the magnetization alignment of contacts A-D \cite{Dery_Nature07}. Examples of the potential level and transient current responses will be discussed in Sec.~\ref{sec:device_sim}. In the original MLG design, semiconductors were employed as the spin transport channel. Here, we propose to use graphene due to its robust room temperature spin transport and high mobility \cite{Tom07,Han09,Han09a,Han10}. Figure~\ref{fig:image} shows an optical microscope image of a lateral graphene sheet topped with ferromagnetic contacts. Note the similarity to the proposed MLG. Non-local spin-valve measurements of this device at room temperature measurements yielded a spin-diffusion length of 3~$\mu$m in the graphene sheet \cite{Han10}.\footnote{In measurements of non-local spin-valve devices, a charge current flows between two biased FM contacts that are spaced by a non-magnetic material (graphene in this case). The generated non-equilibrium spin accumulation in the non-magnetic material is then measured by the change in the potential level of nearby FM contacts outside the region of the charge current flow.} The long spin diffusion length at room temperature is a unique property of graphene that could be further improved with material optimization. The use of tunnel barriers has recently improved the spin polarization of injected carriers to $\sim$30\% \cite{Han10}.

\subsection{MLG Design: Writing Logic Operands and Readout} \label{sec:STT}

Another novelty in the MLG design of Fig.~\ref{fig:mlg} is the spin-transfer torque (STT) magnetization writing \cite{Slonczewski_JMMM96,Berger_PRB96,Katine_PRL00}. Each of the FM contacts employs an all-metallic path with ultra-low resistance for the writing current ($I_w$) during the operand writing phase of the MLG \cite{Braganca_IEEE09,Sun_APL09}. This path allows significant energy savings compared to the standard STT pillar technique based on a magnetic tunnel junction \cite{Huai_APL04,Fuchs_APL04} in which the writing current flows through the MgO tunneling barriers and graphene. In the writing scheme depicted in Fig.~\ref{fig:mlg}, the magnetic moment of the free layer (Py) of a logic operand contact (A-D) is switched by applying a nanosecond-scale current pulse $I_w(t)$ between the two top CoFe layers of the operand contact. The magnetization directions of all CoFe regions are fixed (hard layers). A write current pulse in the opposite direction switches the magnetic moment of the free layer of the operand contact to the opposite orientation.

There are three possible readout schemes for the MLG logic output, all by perturbing the magnetization direction of the free layer in the middle contact, but relying on different configurations of magnetization of the fixed layers of the middle contact. In the first scheme, the fixed layer magnetic moment directions in the middle contact are the same as those of the operand contacts as shown in Fig.~\ref{fig:mlg}. For this configuration, the readout is performed during rotation of magnetization of the free layer in response to a read current pulse $I_r(t)$ in the shape of a single period of a sinusoid. Such a read current pulse of sufficient magnitude switches the magnetic moment of the free layer back and forth and thus results in a rotation of magnetization of the free layer in the plane of the sample by 360$^{\circ}$. In the second readout scheme, the magnetic moments of the fixed layers are rotated clockwise by 90$^{\circ}$ in the plane of the sample compared to the directions shown in Fig.~\ref{fig:mlg}. For this scheme, a positive read current pulse rotates the free layer moment in the sample plane by $\sim$90$^{\circ}$  followed by rotation of magnetization back to the original direction after the read current pulse. The advantage of this scheme is a smaller perturbing read current because perpendicular direction of the fixed layer moment maximizes spin torque magnitude \cite{Slonczewski_JMMM96}. In the third scheme, the fixed layer moments are perpendicular to the sample plane. In this configuration, the free layer moment rotates by 360$^{\circ}$  in response to a short positive read current pulse \cite{Kent_APL04,Nikonov_JAP10}. The fixed layer magnetic configurations for the first two readout schemes can be achieved via a combination of exchange bias and magnetic shape anisotropy \cite{Valenzuela_Nature06}. The magnetic configuration of the third readout scheme can be achieved via the use of magnetic materials with perpendicular magnetic anisotropy \cite{Meng_APL06}. In the following text, we assume the first readout scheme is used, but the other two schemes can be applied without much difficulty.

\subsection{Device Simulation} \label{sec:device_sim}

\begin{figure}
 \center{\includegraphics[width=3in]{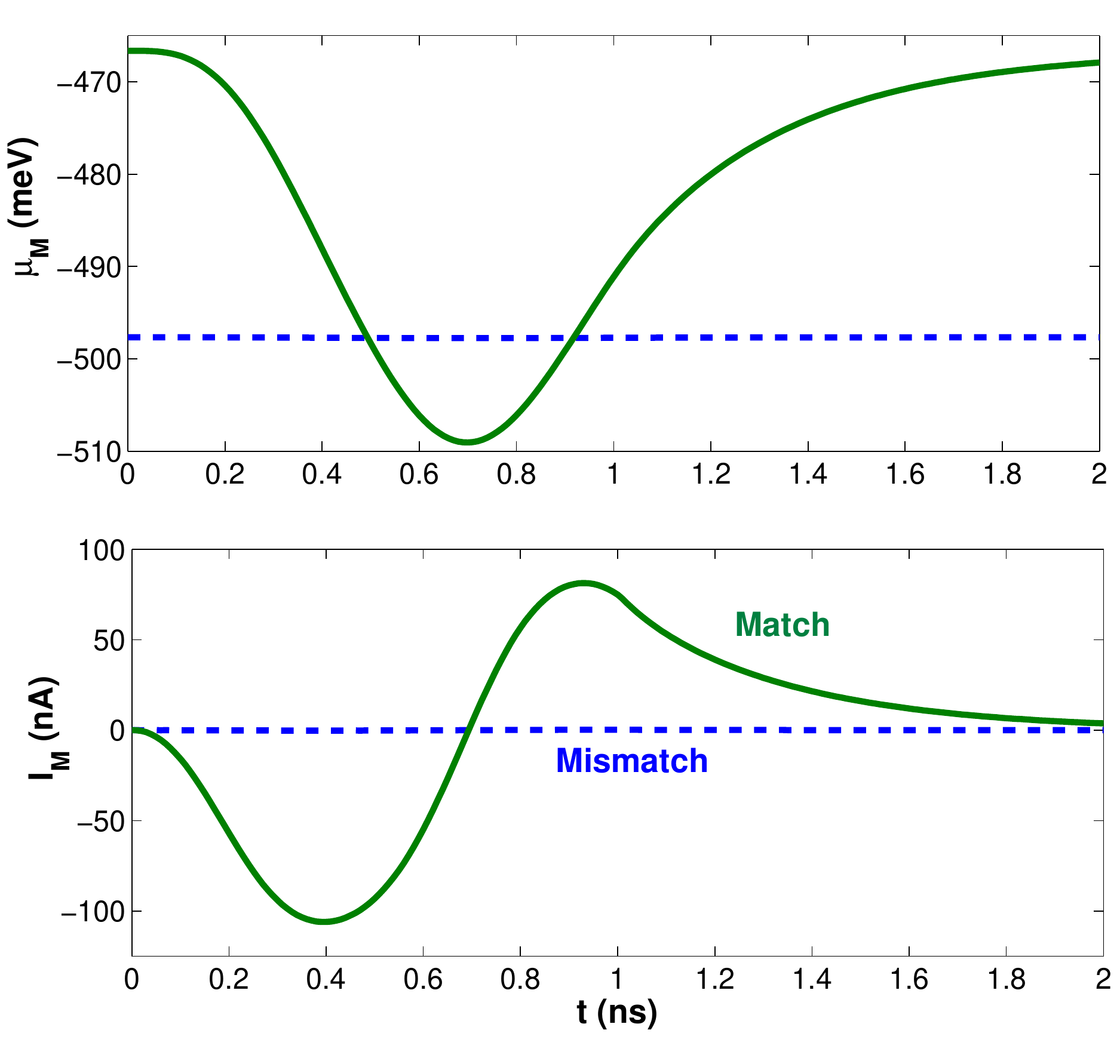}}
 \caption{Modeled electrical behavior of a magneto-logic gate set for matching between the stored (B \& C) and search (A \& D) bits. The upper and lower panels show, respectively, the transient response of the electrochemical potential level in the middle contact and of the transient current across it. The response is to a 1~ns in-plane
rotation of the magnetization direction of M. The bias is V$_{dd}$=1~V and the external capacitor is $C_e$=1~fF. The five contacts are 50~nm wide and 100~nm deep in the z direction (see Fig.~\ref{fig:mlg}). The spacing between contacts is 30~nm. The resistance and intrinsic capacitance of each contact, are respectively, 200~k$\Omega$ and 0.4~fF (areal conductance of 10$^5$~$\Omega^{-1}$cm$^{-2}$ and areal capacitance of 0.08~F/m$^2$).
 \label{fig:mlg_output}}
\end{figure}

The readout operation of a single MLG is simulated during a 1~ns in-plane full rotation of the magnetization of the middle contact. The electrochemical potential level in the  middle contact ($\mu_M$) and the transient current across it ($I_M$) are shown in Fig.~\ref{fig:mlg_output}. In this example the output node (see Fig.~\ref{fig:mlg}) is grounded and the external capacitor is $C_e$=1~fF. If the search key (encoded in both contacts A \& D) match/mismatch the stored bit (encoded in both contacts B \& C) then the transient response is significantly larger/smaller. We duplicate the encoding of the search and stored bits in two contacts each in order to account for the possibility of `don't care' bits (section~\ref{sec:circuit}). The responses in Fig.~\ref{fig:mlg_output} are modeled via a diffusive transport model in the graphene layer which includes the effects of traversing under the finite width of the metal contacts and of the intrinsic capacitance across the tunneling barrier. Complete details of the transport model are given in the supplementary material of \cite{Dery_Nature07}. We have followed the experimental results of Han \textit{et al.} \cite{Han09,Han09a,Han10} and have assumed the
following parameters. The contact polarization is 0.3=$|$R$_\uparrow$-R$_\downarrow$$|$/(R$_\uparrow$+R$_\downarrow$) where R$_{\uparrow(\downarrow)}$ is the contact resistance for majority (minority) spins of the ferromagnet, and R=R$_\uparrow$+R$_\downarrow$ is the contact resistance. The sheet resistance in the graphene layer is 1~k$\Omega$, the spin-diffusion length is 3~$\mu$m, and the diffusion constant is 0.018~m$^2$/s (see Fig.~4(a) in Ref.~\cite{Han10}).

The favorable power scaling with reducing the contact area in STT schemes
\cite{Ralph_JMMM08} is an important benefit of MLGs. The area of an MLG can be
of the order of 0.01~$\mu m^2$ if the area of each STT contact is 50~nm$^2$
\cite{Yoda_CAP10}. Below we simulate a 3.2 Mbits spintronic search engine
with 25K words of 128 bits using 0.04-$\mu$m$^{2}$ MLG cells. In this design,
the total circuit area of the MLGs is less than 1~mm$^{2}$. In comparison, we
find two products with the following area specifications. The \emph{package}
size of a state-of-the-art CMOS based search engine is 729 mm$^{2}$ (Renesas
R8A20410BG; 20 Mbits). Another product from Analog Bits, Inc. provides
512x144bits of CAM in a macro of 1mm$^2$ using TSMC 0.13$\mu m$ technology.
Furthermore, each cell in a CMOS based search engine consumes about 10
transistors and the cell area can be estimated based on actual designs. Based
on the tool provided by Agrawal and Sherwood~\cite{Agrawal_TVLSI08}: such
a cell will be 0.55$\mu m^2$ when implemented with 32~nm CMOS technology.
Thus, we expect spintronic designs of search engines to be orders of
magnitude smaller. Finally, we note that the area of a reprogrammable and
nonvolatile MLG can be further reduced if one uses contacts made of highly
anisotropic magnetic materials for which the thermal stability factor is
higher \cite{Ikeda_IEEE07}.

\section{MLG circuit: spintronic search engine}
\label{sec:circuit}

\begin{figure}
 \center{\includegraphics[width=3.5in]{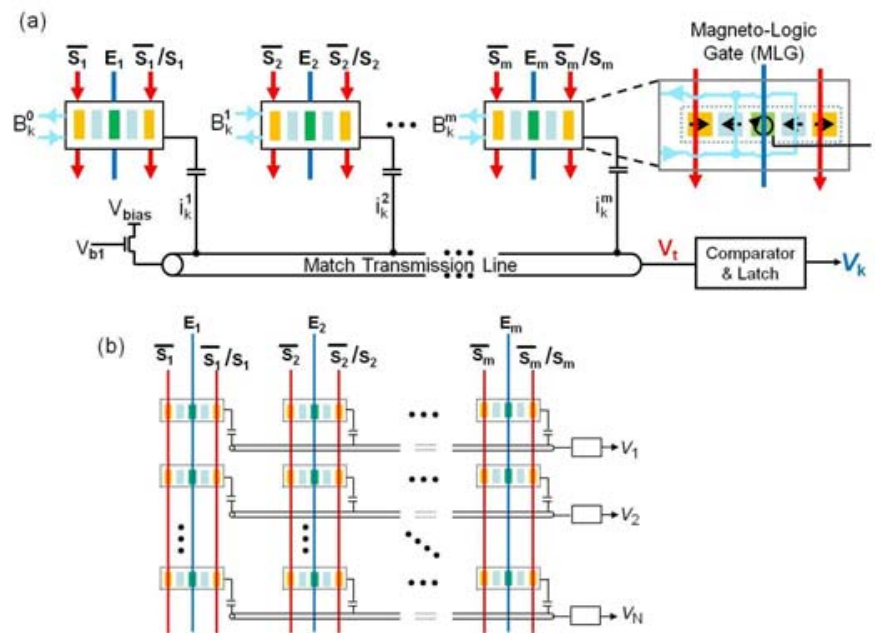}}
 \caption{(a) Spin-based search engine circuitry for one m-bit word using MLG
cells and an AND sensing architecture. Each MLG consists of 5-FM
terminals with an inherent nonvolatile XOR logic arrangement. For don't care bits the
right and left search bit are opposite. The CMOS sensing circuit
compares the match-line voltage output, $V_t$, with a reference
level and generates a digital output voltage, $V_k$, indicating a
match or mismatch in line $k$ (see Fig.~\ref{fig:sense} for
details). (b) Overall search engine architecture (N$\times$m bits).
In the design example, N=25000, m=128.
 \label{fig:circuit}}
\end{figure}

To demonstrate the potential of MLG-based circuits, we use MLGs as
building blocks to construct a spintronic search engine, which
serves as a technology demonstration vehicle. The associative search
of MLGs enables a highly scalable architecture with low power
consumption. Fig.~\ref{fig:circuit}(a) shows circuitry for one m-bit
word in a MLG-based search engine based on STT-MRAM technology. In
each MLG, the current direction in one bit-line (vertical red wire) encodes a
bit of the input search word at the outer contacts. The inner
contacts (not the middle contact) hold the bit of the stored word
(written by the horizontal cyan wires). If the stored and search magnetization
directions are similar/dissimilar, the spin accumulation in the
graphene channel is low/high \cite{Der06a, Der06b}, indicating a
match or mismatch. As discussed in Section~\ref{sec:mlg}, the MLG
carries out a XOR operation between the search bit and stored bit,
and the matching result can be read by perturbing the middle contact
magnetization \cite{Dery_Nature07,Cyw06}. When it is not important
to match a specific bit (Don't-Care bit), the left and right outer
contacts are magnetized antiparallel (`0' and `1'), and the MLG
always outputs a match.

Fig.~\ref{fig:circuit}(b) shows the overall search engine
architecture where multiple m-bit rows are stacked to form an
N$\times$m bit array, where N$\gg$m$\gg$1. In this
configuration, the N MLGs in a column share the same three bit-line
currents which either write the search bit (two red vertical lines
across the outer contacts) or perturb the middle contact (blue
vertical line across the middle contact). Therefore, the current
direction in each of these lines can be controlled by a single
switch circuit. In conventional STT-MRAM designs, on the other hand,
writing to each memory cell requires a dedicated switch circuit to
steer the writing current, which can be as large as over 0.1~mA. To
reduce the power dissipation of the transistors in the switch circuit, most STT-MRAM
designs employ magnetic tunnel junctions since the writing operation
in these devices requires smaller current densities than an
all-metallic STT cell. As a result, the total power consumption of
the transistors and magnetic tunnel junction is reduced in comparison
to that with an all-metallic STT cell. Nonetheless, the power-delay
product of STT-MRAMs is still poor in comparison with conventional
SRAMs (in spite of the improved scaling benefits of STT-MRAMs). The
spin-based search engine design alleviates the power-delay problem
of STT-MRAMs since N$\gg$1 writing operations (e.g., in our design
example N=25,000) are associated with a steering operation of a
single switch circuit. Correspondingly, the MLGs employ all-metallic
paths for the currents during the operand writing phase (contacts
A-D) and the readout phase (contact M). These ultra-low-resistance
paths allow the writing and readout currents to be shared on the
same bit line and hence lead to significant energy savings compared
to the single-cell read/write operation in standard STT-MRAMs.
%for a standard STT cell using a magnetic tunnel junction.

Similarly, the proposed spintronic search engine can achieve a
substantial improvement in power consumption compared to its CMOS
counterparts. In a conventional CMOS search engine, the search bit
is distributed  as charges to all cells on a bit-line, which is
loaded with distributed intrinsic and parasitic capacitance of many
thousands of CMOS memory cells. Hence the fan-out limitation of CMOS
gates limits the number of words. In our spintronic search engine,
the energy consumption of the shared magnetization writing current
is potentially much smaller than what is needed to simultaneously
charge the bit-line in a CMOS implementation. For example, all of the
writing operations in a search engine with 25K words of 128 bits
consume 0.15~W, given that the length of each bit line is 1~mm, the
cross section is 50~nm by 50~nm, the average metal conductivity is
10$^5$~$\Omega^{-1}$cm$^{-1}$, and the current is 0.1~mA. As will be
shown, this power dissipation is a small fraction of the circuit
total power.

%\section{Spintronics-CMOS Interface}
%\label{sec:interface}

One of the key challenges to spintronics lies at the interface
between the spin-based devices and supporting CMOS circuitry.
Specifically, without consuming significant power, we wish to detect
the nA output current pulses from MLGs, and convert them to
rail-to-rail digital signals ($V_k$ in Fig.~\ref{fig:circuit}) for
CMOS electronics.
%\footnote{The index of the matched line is also needed information.
%For example, if the match occurs at the $k^{th}$ line, then $k$ is
%the address in a separated RAM where we can find the port number for
%internet protocol packets.}
%In comparison,
An MLG generates an output current pulse with a small peak amplitude
($\sim$100~nA when matched) and duration ($\sim$1~ns), i.e., a
charge of $\sim$100 electrons. Hence MLGs operate more like small
area DRAM cells but at the speed of larger SRAM cells. The speed and
power consumption of MLG-based search engines, therefore, will be
largely determined by the sensing scheme. In a conventional CMOS
search engine, this task of match-line sensing is typically done by
connecting the cells within a word either in a logic wire-OR
fashion, or as series transmission gates. For MLG-based search
engines, a direct logic wire-OR or transmission-gate structure is
difficult to implement because of the analog nature of the MLG
output, which is a small current pulse. Further, the detection
circuit needs high resolution to resolve the very small difference between
a match case and an all-but-one-bit mismatch case.

%Further, the combined output capacitance of MLGs will limit the
%bandwidth of the sensing circuitry, and in turn limits the number of
%cells on one match-line.

To achieve the required sensitivity and resolution, we propose a
low-noise sensing circuit as shown in Fig.~\ref{fig:sense}. The
outputs of all MLGs are connected to a match-line in parallel, and
the summed current pulses generate a voltage signal at the input of
the detection circuit, which in turn is compared to a reference
voltage to generate a digital output, indicating a match or
mismatch. Low-noise analog sensing significantly improves the
sensitivity compared to a conventional digital sense amplifier. The
power consumption of the search engine is kept low by optimizing
circuit design and operating it at low duty cycle.

\section{A Spintronic Search Engine Design Example}\label{sec:design}

\begin{figure}[htb]
 \center{\includegraphics[width=3.5in]{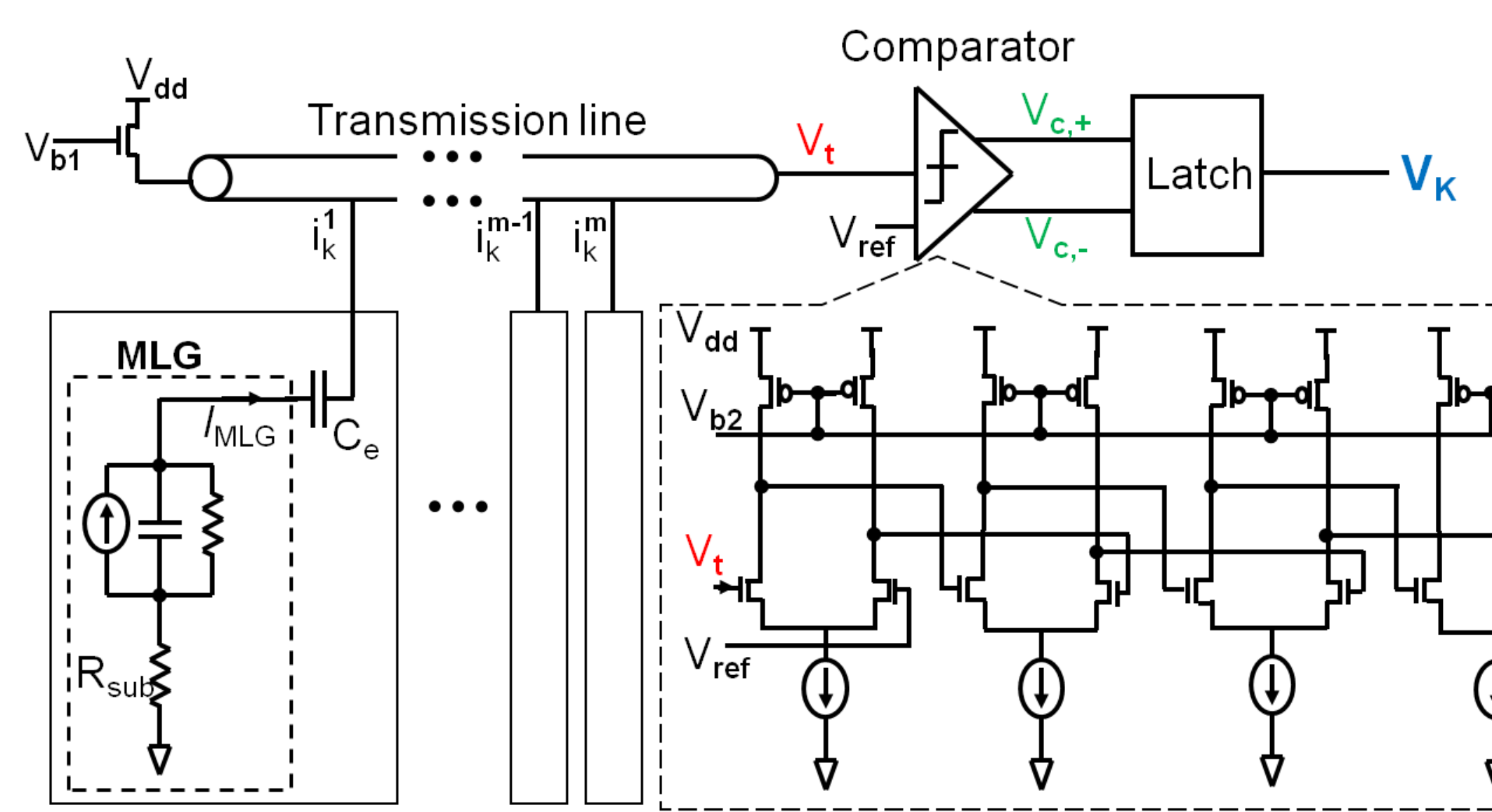}}
 \caption{Circuit schematic for performance evaluation of spintronic
search engine. The MLG model is shown in the dash-line box. The
pulse current source is chosen such that the transient current,
$I_{MLG}$, is similar to the signals of Fig.~\ref{fig:mlg_output} in
both match and mismatch cases. The resistance and capacitance in
parallel to the current source are 200~k$\Omega$ and 0.4~fF,
respectively, representing the intrinsic components of the M contact
in the MLG. The substrate resistance, $R_{sub}$=100~k$\Omega$,
denotes the MLG's distributed resistance to the ground (via the
graphene layer and the grounded B $\&$ C contacts). The coupling
capacitor is $C_e$=1~fF.
 \label{fig:sense}}
\end{figure}

To evaluate the performance of an MLG-based search engine circuit,
we design and simulate the prototype circuit in
Fig.~\ref{fig:sense}. A circuit model is constructed for a
0.04~$\mu$m$^2$ MLG device described in Sec.~\ref{sec:mlg}. The MLG
output port is modeled as a pulse current source, and a parasitic
resistance and capacitance of the M contact. It also includes a
distributed resistance from the M contact to the ground (via the
graphene layer and the grounded contacts, B \& C), shown as
$R_{sub}$. A coupling capacitor $C_e$ is used to isolate the M
contact from the CMOS circuitry in dc, and its value is selected as
a trade-off between signal attenuation and the chip area required to
implement it on-chip. When the output is short-circuit to ground
through $C_e$, this model can reproduce the MLG transient response
in Fig.~\ref{fig:mlg_output}.

The output current pulses of all cells are summed on the
transmission line. Note that this is different from the direct logic
wire-OR structure in CMOS search engines in that each MLG behaves
like a current source instead of a switch, and exhibits a large
impedance to the transmission line in either match or mismatch
cases. The high impedance also helps to reduce the crosstalk between
the MLGs.\footnote{We have also studied an alternative design in
which the MLG output is first amplified by a single-transistor
amplifier to ensure a good signal-to-noise ratio before pulse
combining on the transmission line and to suppress crosstalk. The
power consumption of this alternative sensing circuit is more than
doubled while the (already negligible) crosstalk between MLGs is
further suppressed.} The summed current pulses on the transmission
line generate a voltage signal at the end of the transmission line
($V_t$ in Fig.~\ref{fig:sense}). This voltage signal goes to an
$n$-stage comparator ($n$=4 in this design example). Then a latch
converts the comparator output to a full-swing digital signal,
indicating a match or mismatch. The comparator is optimized for
small offset voltage and large common-mode-rejection-ratio, which
reduce the comparison error. This low-noise analog sensing circuit
exhibits good sensitivity and fine resolution with power consumption
comparable to a conventional sense amplifier in SRAMs.

\begin{figure}
 \center{\includegraphics[width=3.5in]{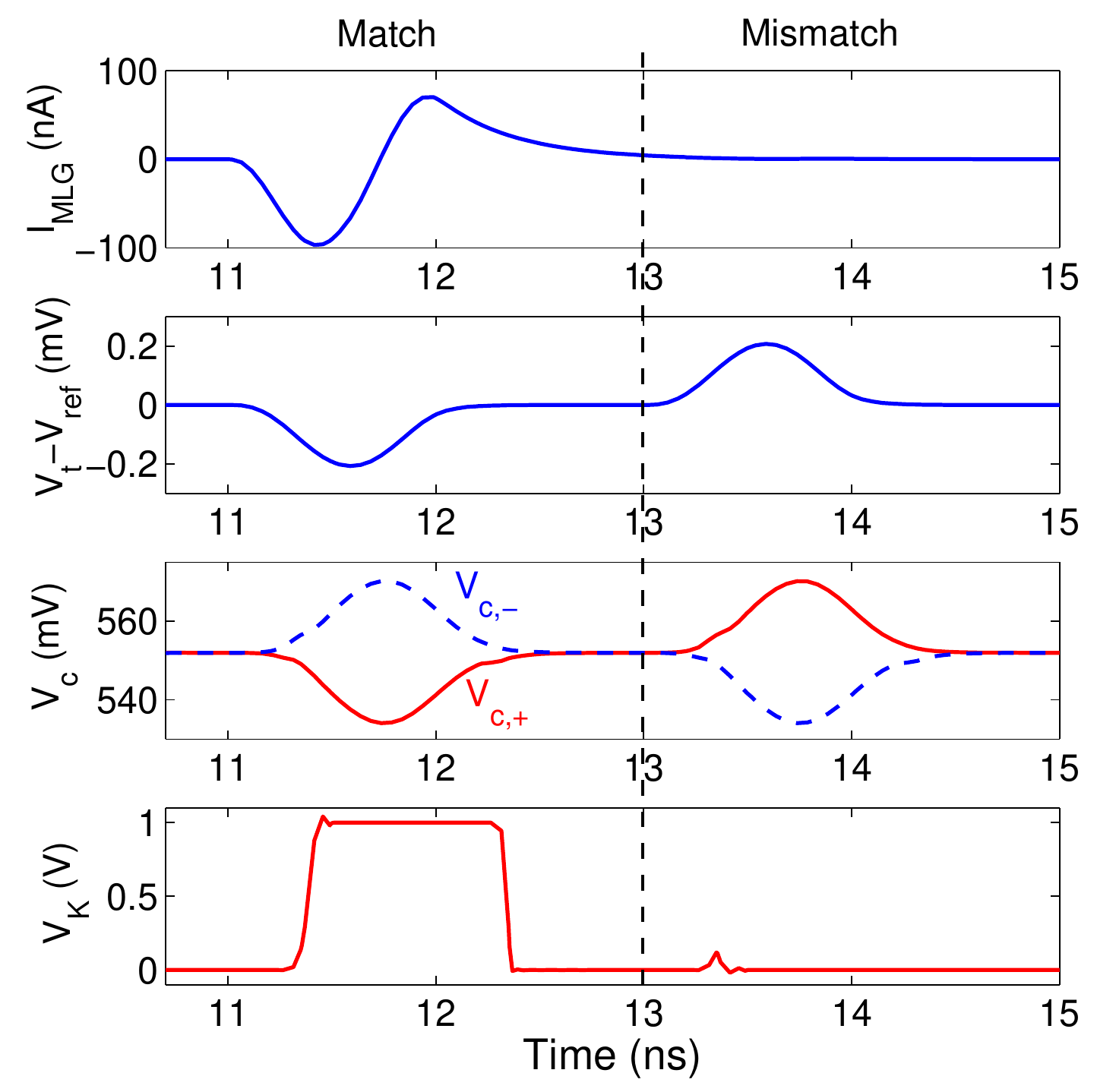}}
 \caption{Signal waveforms for the prototype MLG-based search engine circuitry
in Fig.~\ref{fig:circuit} from ADS simulation. All graphs show a
match case (between 11 ns and 13 ns) followed by the worse case
mismatch scenario. The former (latter) correspond to cases where 128
(127) out of 128 bits match. (a) Current signal at the MLG output
prior to the external capacitor. (b) Difference between the
generated transmission line voltage and the reference level,
$V_t-V_{ref}$, at the input for the sensing circuit. (c) Voltage
output of a 4-stage comparator, $V_C=V_{C,+}-V_{C,-}$. (d) Digital
voltage output of the latch, $V_K$.
 \label{fig:signals}}
\end{figure}

The prototype circuit is simulated using a high speed circuit
simulator, Advanced Design System (ADS). The CMOS transistors are
based on 45-nm Predictive Technology Model \cite{Zha06}, and the
power supply voltage is $V_{dd}$=1~V. Fig.~\ref{fig:signals} shows
the output signal waveforms when the sensing circuit operates at 500
MHz with approximately 50\% duty cycle. Fig.~\ref{fig:signals}(a)
shows the current signal at the output of the MLG which replicates
the match and mismatch signal behaviors in
Fig.~\ref{fig:mlg_output}. Fig.~\ref{fig:signals}(b) shows the value
of $V_t-V_{ref}$ at the input for the sensing circuit. $V_t$ is the
transmission line voltage output and $V_{ref}$ simulates the worst
case scenario in which 127 out of 128 bits are matched and the
current signal of one bit is halfway between a match and a mismatch.
Note the opposite polarity of the signal between a match and a
mismatch cases. Fig.~\ref{fig:signals}(c) shows the output of the
4-stage comparator, $V_C=V_{C,+}-V_{C,-}$. In all mismatch
scenarios, the comparator output is always positive ($V_C>0$). The
positive amplitude increases with the number of mismatched bits.
Only with a line match $V_{C}$ is negative and is of the order of
-20~mV. Fig.~\ref{fig:signals}(d) shows the digital output of the
latch.

It is evident that the overall gain is evenly distributed among the
stages to achieve the best signal-to-noise ratio and maintain the
large bandwidth. The 128-bit match-line sensing circuit can operate
up to 500~MHz, and detect MLG output current difference down to
9~nA.
%from 118~nA to 23.7~$\mu$A.
The 4-stage comparator and the latch use 0.26~mW. For a 3.2-Mbit search engine
(25K words of 128 bits), the total power consumption
of these CMOS circuitry is 6.5~W. The power dissipation of each MLG
is mostly due to the DC current that flow between the A and B
contacts as well as the C and D contacts. In our simulated
200~K$\Omega$ contacts and V$_{dd}$=1~V applied bias, this
corresponds to 5~$\mu$W per MLG (16~W for 3.2~Mbit search engine).
Together with the (much smaller) power consumption of all latches in
the sensing circuits and the STT-writing operations, the total power
consumption of the MLG-based search engine is about 23~W. In spite of
the fact that we have not employed optimization techniques (e.g.,
pipelined search) this power is already lower compared to optimized
CMOS counterpart designs. In addition, using tunneling barriers in
the MLG that are more resistive and more spin polarized will keep
the MLG current output at the same magnitude but will use less
power. For example, if smaller MLG contacts with resistance of R=R$_\uparrow$+R$_\downarrow$=2000~k$\Omega$ and polarization of 0.9=$|$R$_\uparrow$-R$_\downarrow$$|$/R can be fabricated then the transient responses are kept similar to those in Fig.~\ref{fig:mlg_output} but the MLG power dissipation drops by an order of magnitude (from 5~$\mu$W to 0.5~$\mu$W per MLG and from 23~W to 9~W for the entire spintronic-CMOS circuit).

%The energy efficiency of the search engine is  corresponding to an
%energy efficiency of about 10~fJ per bit search operation.

The match-line power consumption depends on the MLG output signal
amplitude. The larger the output signal, the less gain needed for
amplification, and hence less power consumption. In our simulations
we have assumed 200:1 MLG output current ratio for match vs.
mismatch. The power consumption remains comparable for 10:1
simulated ratio with slight modification of the circuit parameters.
It is expected that as the MLG performance improves
(larger output signal amplitude, smaller parasitic capacitance and
less device variation), the energy efficiency will increase even
more since larger number of bits can share the match-line.
Correspondingly, the CMOS sensing circuitry needs to improve its
resolution.

The circuit area of the sensing circuit is fairly small, and the
area overhead is negligible since it is mortgaged by the large
number of bits per search word. For example, the aforementioned
comparator and latch use a chip area of less than 3~$\mu$m$^2$ per
match-line. In comparison, the conventional CMOS search engine needs
a substantial area overhead due to the much shorter search word.
Another important issue for a spin-based search engine is the MLG
parameter variations. These variations are often being overlooked
and their consideration is paramount to any integrated circuit.
Without corresponding circuit provisions, the output current
variation would fundamentally limit the length of search words
(e.g., the fidelity of 128 bits word is the same as of a 128, 87,
and 9 bit strings if the MLG output variations are, respectively,
0.1\%, 1\% and 10\%). One way to alleviate the variation concern and
yet to achieve longer words is by adopting an offset cancelation
technique, i.e., to first calibrate each match-line, and then remove
the mismatch between them during the evaluation.

\section{Potential Applications} \label{sec:apps}
In addition to non-volatility, the MLG circuit enables a range  of associative computing applications, whereby a large amount of information can be accessed and filtered in parallel with energy efficiency and high speed. As mentioned above, efficient search capabilities are clearly useful in classical examples such as network routing tables. But more importantly, these capabilities can critically impact more complex applications that are limited by the serial communication and processing bottleneck sometimes referred to
as the von Neumann bottleneck. Leading examples of applications include the increasingly widely used business analytics and automatic fraud detection that pore over huge amount of data searching for insights. Because of the growing importance of such applications, conventional optimizations are being vigorously pursued in the research community and in commercial products. However, these approaches can reduce some bandwidth wastes but still fundamentally require complete scanning that subjects the large data arrays through the von Neumann bottleneck. In contrast, the underlying MLG circuit allows us to support such applications with a new processing substrate that allows in-memory, massively parallel
processing of data and thus enable orders-of-magnitude improvement in processing speed. As an example, conventional business analytics systems are bottlenecked by the sequential  scanning of large amount of data that cannot fit on-chip caches. The scanning operations are almost universal simple such as a string and integer comparison. Invariably, such searches return exceedingly small portions of the data (e.g., $<$0.1\%) as results for further processing. MLG search circuit embedded with the storage element thus allows irrelevant data to be filtered out in parallel with essentially constant time and reduce the communication need by orders of magnitude and thus completely eliminate the von Neumann bottleneck.

The logic operations as performed by MLGs also represent a potentially large technology leap for space vehicle processing.  Most importantly, it is because the logic operations appear to be inherently radiation hardened.  For example, due to the manner in which the logic is stored and read, single event effects should no longer cause upset operations or failures, as could occur in a current field effect transistor.  Additionally, the non-volatile logic operations are preserved in the event that systems need to be powered down to protect the electronics.  This aspect results in faster times for bringing such systems back online.  Finally, the scalability of this circuitry coupled with the reduced power consumption (as compared to current CMOS technology) appears to fit the tight constraints associated with space vehicle electronics designs, opening the door to potential on-board processing.

\section{Conclusion} \label{sec:conc}
The performance of highly-scalable MLGs will offer much faster operation speed, smaller area, and better energy efficiency than their CMOS counterparts. It is expected that in  5-10 years, the MLG operation speed can reach 0.1~ns, while consuming 0.01~$\mu$m$^2$ chip area and 100~aJ/operation. It is also envisioned that graphene can be utilized to replace silicon and achieve even greater performance at room temperature. The application of spin-based devices \cite{Dery_Nature07,Zutic_RMP04} in nonvolatile logic circuits will represent a disruptive advance in the design and implementation of some of the most critical building blocks in general-purpose microprocessors and other high-performance  computing and communication systems. It will lead to significant performance gains and energy savings in such systems. Fundamentally, it will enable a paradigm change in computer  architectures from von Neumann to one in which memory and processing are seamlessly integrated together.

\section*{Acknowledgement}
We deeply thank Jiangyun Hu and Jie Zhang for their help in the circuit design and simulations and Ovunc Kocabas for his help in analyzing conventional CAM designs.


\begin{thebibliography}{99}

\bibitem{Dery_Nature07}
H. Dery, P. Dalal, L. Cywinski, and L. J. Sham, ``Spin-Based Logic in Semiconductors for Reconfigurable Large-Scale Circuits'', \textit{Nature}, vol. 447, no. 7144, pp. 573-576, May 2007.

\bibitem{Kom93} E. Komoto, T. Homma, and T. Nakamura, ``A high-speed and compactsize JPEG Huffman decoder using CAM'', in \textit{Symp. VLSI Circuits Dig. Tech. Papers}, 1993, pp. 37-38.

\bibitem{Wei93} B. W. Wei, R. Tarver, J.-S. Kim, and K. Ng, ``A single chip Lempel-Ziv data compressor'', in \textit{Proc. IEEE Int. Symp. Circuits Syst. (ISCAS)}, 1993, pp. 1953-1955.

\bibitem{Yan94} R.-Y. Yang and C.-Y. Lee, ``High-throughput data compressor designs using content addressable memory'', in \textit{Proc. IEEE Int. Symp. Circuits Syst. (ISCAS)}, 1994, pp. 147-150.

\bibitem{Cra98} D. J. Craft, ``A fast hardware data compression algorithm and some algorithmic extensions'', \textit{IBM J. Res. Devel.}, vol. 42, no. 6, pp. 733-746, Nov. 1998.

\bibitem{Liu94} L.-Y. Liu, J.-F.Wang, R.-J.Wang, and J.-Y. Lee, ``CAM-based VLSI architectures for dynamic Huffman coding'', \textit{IEEE Trans. Consumer Electron.}, vol. 40, no. 3, pp. 282-289, Aug. 1994.

\bibitem{Mer00} M. Meribout, T. Ogura, and M. Nakanishi, ``On using the CAM concept for parametric curve extraction'', \textit{IEEE Trans. Image Process.}, vol. 9, no. 12, pp. 2126-2130, Aug. 2002.

\bibitem{Nak00} M. Nakanishi and T. Ogura, ``Real-time CAM-based Hough transform and its performance evaluation'', \textit{Machine Vision Appl.}, vol. 12, no. 2, pp. 59-68, Aug. 2000.

\bibitem{ITRS09} ITRS 2009. [Online]. Available: \url{http://www.itrs.net/links/2009ITRS/Home2009.htm}

\bibitem{Rakheja10} S. Rakheja and A. Naeemi, ``Interconnects for Novel State Variables: Performance Modeling and Device and Circuit Implications'', \textit{IEEE Trans. Electron Devices}, vol. 57, no. 10, pp. 2711-2718 , Oct. 2010.

\bibitem{Zutic_RMP04}  I. \v{Z}uti\'{c}, J. Fabian, and S. Das Sarma, ``Spintronics: Fundamentals and applications'', \textit{Rev. Mod. Phys.}, vol. 76, no. 2, pp. 323-410,  Apr. 2004.

\bibitem{Tom07} N. Tombros, C. Jozsa, M. Popinciuc, H. T. Jonkman, and B. J. van Wees,  ``Electronic spin transport and spin precession in single graphene layers at room temperature'', \textit{Nature} 448, no. 7153, pp. 571-574, Aug. 2007.

\bibitem{Han09} W. Han, K. Pi, W. Bao, K. M. McCreary, Y. Li, W. H. Wang, C. N. Lau, and  R. K. Kawakami, ``Electrical detection of spin precession in single layer graphene spin valves with  transparent contacts'', \textit{Appl. Phys. Lett.}, vol. 94, no. 22, p. 222109, Jun. 2009.

\bibitem{Han09a} W. Han, W. H. Wang, K. Pi, K. M. McCreary, W. Bao, Y. Li, F. Miao, C. N. Lau, and R. K. Kawakami, ``Electron-hole asymmetry of spin injection and transport in single-layer graphene'', \textit{Phys. Rev. Lett.}, vol. 102, no. 13, p. 137205, Apr. 2009.

\bibitem{Han10}  W. Han, K. Pi, K. M. McCreary, Y. Li, Jared J. I. Wong, A. G. Swartz, and R. K. Kawakami, ``Tunneling Spin Injection into Single Layer Graphene'', \textit{Phys. Rev. Lett.}, vol. 105, no. 16, p. 167202, Oct. 2010.

\bibitem{Slonczewski_JMMM96} J. C. Slonczewski, ``Current-driven excitation of magnetic multilayers'', \textit{J. Magn. Magn. Mater.}, vol. 159, no. 1-2, pp. L1-L7,  Jun. 1996.

\bibitem{Berger_PRB96} L. Berger, ``Emission of spin waves by a magnetic multilayer traversed by a current'', \textit{Phys. Rev. B}, vol. 54, no. 13 , pp. 9353-9358, Oct. 1996.

\bibitem{Katine_PRL00} J. A. Katine, F. J. Albert, R. A. Buhrman, E. B. Myers, and D. C. Ralph, ``Current-driven magnetization reversal and spin-wave excitations in Co/Cu/Co pillars'', \textit{Phys. Rev. Lett.}, vol. 84, no. 14, pp. 3149-3152, Apr. 2000.

\bibitem{Braganca_IEEE09} P. M. Braganca, J. A. Katine, N. C. Emley, D. Mauri, J. R. Childress, P. M. Rice, E. Delenia, D. C. Ralph,  R. A. Buhrman, ``A Three-Terminal Approach to Developing Spin-Torque Written Magnetic Random Access Memory Cells'', \textit{IEEE Trans. Nanotech.}, vol. 8, no. 2, pp. 190-195, Mar. 2009.

\bibitem{Sun_APL09}  J. Z. Sun, M. C. Gaidis, E. J. O'Sullivan, E. A. Joseph, G. Hu, D. W. Abraham, J. J. Nowak, P. L. Trouilloud, Yu Lu, S. L. Brown, D. C. Worledge, and W. J. Gallagher, ``A three-terminal spin-torque-driven magnetic switch'', \textit{Appl. Phys. Lett.}, vol. 95, no. 8, p. 083506, Jul. 2009.

\bibitem{Huai_APL04} Y. Huai, F. Albert, P. Nguyen, M. Pakala, and T. Valet, ``Observation of spin-transfer switching in deep submicron-sized and low-resistance magnetic tunnel junctions'', \textit{Appl. Phys. Lett.}, vol. 84, no. 16, pp. 3118-3120, Feb. 2004.

\bibitem{Fuchs_APL04} G. D. Fuchs, N. C. Emley, I. N. Krivorotov, P. M. Braganca, E. M. Ryan, S. I. Kiselev, J. C. Sankey, D. C. Ralph, R. A. Buhrman, and J. A. Katine, ``Spin-Transfer Effects in Nanoscale Magnetic Tunnel Junctions'', \textit{Appl. Phys. Lett.}, vol. 85, no. 7,  pp. 1205-1207, Jun. 2004.

\bibitem{Kent_APL04}  A. D. Kent, B. Ozyilmaz, E. del Barco, ``Spin-transfer-induced precessional magnetization reversal'', \textit{Appl. Phys. Lett.}, vol. 84, no. 19, pp. 3897-3899, May 2004.

\bibitem{Nikonov_JAP10}  D. E. Nikonov, G. I. Bourianoff, G. Rowlands, I. N. Krivorotov, ``Strategies and tolerances of spin transfer torque switching'', \textit{J. Appl. Phys.}, vol. 107, no. 11, p. 113910, Jun. 2010.

\bibitem{Valenzuela_Nature06}  S. O. Valenzuela, M. Tinkham, ``Direct electronic measurement of the spin Hall effect'', \textit{Nature}, vol. 442, no. 7099, pp. 176-179, Jul. 2006.

\bibitem{Meng_APL06} H. Meng, J. P. Wang, ``Spin transfer in nanomagnetic devices with perpendicular anisotropy'', \textit{Appl. Phys. Lett.}, vol. 88, no. 17, p. 172506, Apr. 2006.

\bibitem{Ralph_JMMM08} D. C. Ralph and M. D. Stiles, ``Spin Transfer Torques'', \textit{J. Magn. Magn. Mater.}, vol. 320, no. 7, pp. 1190-1216, Apr. 2008.

\bibitem{Yoda_CAP10} H. Yoda \textit{et al.}, ``High efficient spin transfer torque writing on perpendicular magnetic tunnel junctions for high density MRAMs'', \textit{Curr. Appl. Phys.}, vol. 10, no. 1, supplement 1, pp. e87-e89, Jan. 2010.
%  T. Kishi, T. Nagase, M. Yoshikawa, K. Nishiyama, E. Kitagawa, T. Daibou, M. Amano, N. Shimomura, S. Takahashi, T. Kai, M. Nakayama, H. Aikawa, S. Ikegawa, M. Nagamine, J. Ozeki, S. Mizukami, M. Oogane, Y. Ando, S. Yuasa, K. Yakushiji, H. Kubota, Y. Suzuki, Y. Nakatani, T. Miyazaki, and K. Ando,

\bibitem{Agrawal_TVLSI08} B. Agrawal and T. Sherwood, ``Ternary CAM Power and Delay Model: Extensions and Uses'',  \textit{IEEE Transactions on VLSI Systems}, vol. 16, no.5, pp. 554-564, May 2008.

\bibitem{Ikeda_IEEE07} S. Ikeda, J. Hayakawa, Y. M. Lee, F. Matsukura, Y. Ohno, T. Hanyu, and H. Ohno, ``Magnetic Tunnel Junctions for Spintronic Memories and Beyond'', \textit{IEEE Trans. Electron Devices}, vol. 54, no. 5, pp. 991-1002 , Nov. 2007.

\bibitem{Der06a} H. Dery, L. Cywinski, and L. J. Sham, ``Lateral Diffusive Spin Transport in Layered Structures'', \textit{Phys. Rev. B.}, vol.  73, no. 4, p. 041306(R), Jan. 2006.

\bibitem{Der06b} H. Dery, L. Cywinski, and L. J. Sham, ``Spin Transference and Magnetoresistance Amplification in a Transistor'', \textit{Phys. Rev. B.}, vol.  73, no. 16, p. 161307(R), Apr. 2006.

\bibitem{Cyw06} L. Cywinski, H. Dery, and L. J. Sham, ``Electric Readout of Magnetization Dynamics in a Ferromagnet-Semiconductor System'', \textit{Appl. Phys. Lett.}, vol.  89, no. 4, p. 042105, Jul. 2006.

\bibitem{Zha06} W. Zhao and Y. Cao, ``New Generation of Predictive Technology Model for Sub-45 nm Early Design Exploration'', \textit{IEEE Trans. Electron Devices}, vol. 53, no. 11, pp. 2816-2823 , Nov. 2006.


\end{thebibliography}
\end{document}